\documentclass[times,twocolumn,final,authoryear]{elsarticle}

\usepackage{prletters}
\usepackage{framed,multirow}

\usepackage{amssymb}
\usepackage{latexsym}

\usepackage{url}
\usepackage{xcolor}
\definecolor{newcolor}{rgb}{.8,.349,.1}

\usepackage{acronym}
\usepackage{amsmath}
\usepackage{algorithm2e}
\usepackage{arydshln}
\usepackage[utf8]{inputenc}

\journal{Pattern Recognition Letters}

\begin{document}

\newacro{AvS}{Average Similarity}
\newacro{DTW}{Dynamic Time Warping}
\newacro{FFT}{Fast Fourier Transform}
\newacro{GMM}{Gaussian Mixture Model}
\newacro{GRASSHOPPER}{Graph Random-walk with Absorbing StateS that HOPs
among PEaks for Ranking}
\newacro{GSV}{Gaussian Super Vector}
\newacro{HMM}{Hidden Markov Model}
\newacro{ISMIR}{International Society for Music Information Retrieval}
\newacro{KL}{Kullback-Leibler}
\newacro{LSA}{Latent Semantic Analysis}
\newacro{MER}{Music Emotion Recognition}
\newacro{MFCC}{Mel Frequency Cepstral Coefficient}
\newacro{MIREX}{Music Information Retrieval Evaluation eXchange}
\newacro{MIR}{Music Information Retrieval}
\newacro{MMR}{Maximal Marginal Relevance}
\newacro{MSE}{Mean Squared Error}
\newacro{PCA}{Principal Component Analysis}
\newacro{pp}{percentage points}
\newacro{RMS}{Root Mean Square}
\newacro{SDT}{Sound Description Toolbox}
\newacro{SGM}{Single Gaussian Model}
\newacro{SONE}{Specific Loudness Sensation Coefficients}
\newacro{SVD}{Singular Value Decomposition}
\newacro{SVM}{Support Vector Machine}
\newacro{SVR}{Support Vector Regressor}

\begin{frontmatter}

\title{An information-theoretic approach to machine-oriented music summarization}

\author[1,3]{Francisco Afonso \snm{Raposo}\corref{cor1}}
\cortext[cor1]{Corresponding author:
  Tel.: +351-919-360-604}
\ead{francisco.afonso.raposo@tecnico.ulisboa.pt}
\author[1,3]{David Martins \snm{de Matos}}
\author[2,3]{Ricardo \snm{Ribeiro}}

\address[1]{Instituto Superior Técnico, Universidade de Lisboa, Av. Rovisco Pais, 1049-001 Lisboa, Portugal}
\address[2]{Instituto Universitário de Lisboa (ISCTE-IUL), Av. das Forças Armadas, 1649-026 Lisboa, Portugal}
\address[3]{INESC-ID Lisboa, R. Alves Redol 9, 1000-029 Lisboa, Portugal}

\received{1 May 2013}
\finalform{10 May 2013}
\accepted{13 May 2013}
\availableonline{15 May 2013}
\communicated{S. Sarkar}

\begin{abstract}
Music summarization allows for higher efficiency in processing, storage, and sharing of datasets. Machine-oriented approaches, being agnostic to human consumption, optimize these aspects even further. Such summaries have already been successfully validated in some \acs{MIR} tasks. We now generalize previous conclusions by evaluating the impact of generic summarization of music from a probabilistic perspective. We estimate Gaussian distributions for original and summarized songs and compute their relative entropy, in order to measure information loss incurred by summarization. Our results suggest that relative entropy is a good predictor of summarization performance in the context of tasks relying on a bag-of-features model. Based on this observation, we further propose a straightforward yet expressive summarizer, which minimizes relative entropy with respect to the original song, that objectively outperforms previous methods and is better suited to avoid potential copyright issues.
\end{abstract}


\end{frontmatter}


\section{Introduction}

Music summarization is an important task in \ac{MIR} which can be categorized into two types: human- and machine-oriented. Human-oriented summarization needs to take into account that humans will consume the summary of the creative artifact \citep{Bartsch2005,Cooper2002,Cooper2003,Chai2006,Chu2000,Glaczynski2011,Peeters2002,Peeters2003}. Thus, perceptually relevant requirements are at play, such as clarity and coherence, so that people can enjoy listening to the audio summaries. Machine-oriented summarization, however, is agnostic to the fact that it deals with creative artifacts, that is, its purpose is to output a shorter version of the song whose content is optimal, e.g., in terms of relevance and diversity. Generic media-agnostic summarizers have been originally developed for text and transcribed speech \citep{Carbonell1998,Erkan2004,Landauer1997,Mihalcea2004,Ribeiro2011,Zhu2007} and were also applied to video through subtitles and scripts \citep{Aparicio2016}. These methods have no explicit criteria for taking clarity or coherence into account, yielding undesirable effects in the music summaries (for human consumption), e.g., harsh discontinuities or lack of beat synchronization. However, these summaries are useful for automatic (and optimized) music processing, namely, for genre classification tasks \citep{Raposo2015,Raposo2016}.

In this work, we approach machine-oriented summarization from an information-theoretic perspective: we assess its performance, in a task-agnostic way, by measuring the similarity of probabilistic descriptions of music. We estimate single Gaussian distributions (\acp{SGM}) of audio descriptions, which provide a high-level, probabilistic explanation of the data. Then, we use the \ac{KL} divergence as a measure of how well an \ac{SGM} (representing a summary) represents the original data (best represented by its own \ac{SGM}), which reflects how well a summary clip represents the original song. Thus, we can generically assess the information content of summarized music by simply measuring information loss. The \ac{KL} divergence has already been applied this way in text summarization by \cite{Louis2013}. Moreover, we propose a simple yet expressive method, focusing on minimizing the \ac{KL} divergence between the original and summarized songs, that objectively outperforms previous state-of-the-art methods and whose summaries are less prone to infringe copyrights. Benefits of using such summarized data include faster processing, less disk space use, more efficient use of bandwidth, and alleviation of potential copyright issues.

We correlate the results of task-agnostic evaluations with proxy task evaluations (i.e., genre classification and emotion regression), validating the use of the \ac{KL} divergence. We experiment with 2 datasets, each designed for its own proxy task, and the following summarizers: \acs{GRASSHOPPER}, LexRank, \acs{LSA}, \acs{MMR}, and Support Sets. We also summarize using \ac{AvS} and fixed-segments, as continuous baselines, and our proposed Gaussian sampler summarizer. Furthermore, we show that generic summarizers outperform the baselines and that our proposed method outperforms all others, according to all evaluations. These results strengthen and generalize previous conclusions derived by \cite{Raposo2015,Raposo2016}.

The rest of the article is organized as follows: section \ref{sec:previous-work} overviews previous work. In section \ref{sec:gaussian-sampler}, we describe our proposed method. The \ac{KL} divergence is described in section \ref{sec:kullback-leibler}. Section \ref{sec:experiments} details the experiments. Section \ref{sec:results} reports results on the performance of summaries. Section \ref{sec:discussion} discusses the results, and Section \ref{sec:conclusions} concludes the paper and considers future work.

\section{Previous work}
\label{sec:previous-work}

The focus of human-oriented music summarization is on extracting an enjoyable summary that people can listen to clearly and coherently. Since we consider summaries exclusively for automatic consumption, music-specific algorithms, as well as many of their issues and requirements, are outside the scope of this paper. A full discussion concerning this and the usage of other audio proxies (such as \acp{GMM}) can be found in \cite{Raposo2016}. We employ the following media-agnostic summarizers: \ac{GRASSHOPPER} \citep{Zhu2007}, LexRank \citep{Erkan2004}, \ac{LSA} \citep{Landauer1997}, \ac{MMR} \citep{Carbonell1998}, and Support Sets \citep{Ribeiro2011}. We also consider a music-specific summarizer, \ac{AvS} \citep{Cooper2002}, as a continuous baseline.

Music summarization for machine consumption has already been evaluated in the past by \cite{Raposo2015,Raposo2016}, in the contexts of binary and multiclass music genre classification. In those instances, generic summarization methods, which were originally developed for text and speech summarization, were proven to be very effective at machine-oriented music summarization. However, the application of these algorithms to music is not straightforward, since an initial fixed-size segmentation of the songs and a discretization step must be performed in order to map the continuous stream of real-valued audio features to the discrete concepts of phrases and terms (e.g., tf-idf vectors), respectively. This must be done for all of these generic summarizers, since all of them build summaries by ranking phrases and picking the top ranked ones until reaching the required summary length. In turn, the resulting audio summaries are characterized by a concatenation of continuous audio phrases. For instance, if we consider 0.5s terms, 10-term phrases, and a 30s summary, these summarizers output a summary consisting of a concatenation of six 5s phrases. The difference between these algorithms lies solely in the way they rank phrases. Despite being well-suited for machine-oriented tasks, the process leads to summaries that are not enjoyable to humans, since there usually exist harsh discontinuities between phrases, even though each phrase is a continuous segment. An overview of these methods can be found in \cite{Raposo2016}.

\section{Gaussian sampler summarization}
\label{sec:gaussian-sampler}

In this section, we propose a novel method for machine-oriented summarization that aims at building summaries whose Gaussian distribution of the data is as close as possible to the Gaussian distribution of the data of original song. Note that this corresponds to building a summary in a way that asymptotically minimizes the \ac{KL} divergence between both distributions, that is, building a summary that minimizes information loss.

The summarization procedure consists in estimating a multivariate \ac{SGM} for the original song and, iteratively, drawing synthetic samples from that distribution and picking the closest frame to the sample, using the scale-invariant Mahalanobis distance \citep{Mahalanobis1936}. We are essentially sampling from the original pool of frames. However, the frames in the audio are rarely exactly equal to the generated samples. Therefore, the distribution of the selected frames will not be exactly equal to the distribution of the samples, namely, their mean will be shifted. In order to minimize this error, we introduce a heuristic that updates a difference vector (initialized as 0), keeping track of the resulting shift and influencing frame selection in every iteration. The following pseudo-code illustrates this procedure:

\begin{algorithm}
\small
\SetKwInOut{Input}{input}
\SetKwInOut{Output}{output}
\SetKwFunction{gaussian}{gaussian}
\SetKwFunction{mahalanobisargmin}{mahalanobis\_argmin}
\SetKwFunction{size}{size}
\SetKwFunction{samplef}{sample}
\SetKwData{originalframes}{frames}
\SetKwData{n}{n}
\SetKwData{summaryframes}{summary}
\SetKwData{originalgaussian}{sgm}
\SetKwData{diffvector}{diff}
\SetKwData{sample}{sample}
\SetKwData{frame}{frame}
\Input{A set \originalframes of feature vectors; summary size \n}
\Output{A set \summaryframes of feature vectors}
\BlankLine
$\originalgaussian \leftarrow \gaussian{\originalframes}$\;
$\summaryframes \leftarrow \emptyset$\;
$\diffvector \leftarrow 0$\;
\While{$\size{\summaryframes} < \n$}{
  $\sample \leftarrow \samplef{\originalgaussian} - \diffvector$\;
  $\frame \leftarrow \mahalanobisargmin{\sample, \originalframes}$\;
  $\diffvector \leftarrow \sample - \frame$\;
  $\summaryframes \leftarrow \summaryframes \cup \{\frame\}$\;
  $\originalframes \leftarrow \originalframes \setminus \{\frame\}$\;
}
\normalsize
\caption{Gaussian sampler summarizer}
\label{alg:sampler}
\end{algorithm}

Note that this greedy algorithm does not guarantee a global minimum of the \ac{KL} divergence. However, it is simple to implement and effective as validated by the experimental results.

\section{\ac{KL} divergence}
\label{sec:kullback-leibler}

In this work, we generalize conclusions of previous work by evaluating machine-oriented summarization in an information-theoretic and task-agnostic way, i.e., by computing information loss according to the \ac{KL} divergence \citep{Kullback1951}. The \ac{KL} divergence, also called relative entropy, is a non-symmetric measure of the difference between two probability distributions $p$ and $q$. Specifically, the \ac{KL} divergence of $q$ from $p$, $\operatorname{D}_{\text{KL}}\left(p||q\right)$, is a measure of information gain achieved by using $p$ instead of $q$. In other words, it is a measure of how much information is lost when $q$ is used as an approximation of $p$. In this work, $q$ always models some summarized version of the original data, which itself is modeled by $p$, so relative entropy is ideal for measuring how much information is lost by the corresponding summarization process. The \ac{KL} divergence between two Gaussians $\mathcal{N}_0$ and $\mathcal{N}_1$ is defined as:
\begin{multline}
\operatorname{D}_{\text{KL}}\left(\mathcal{N}_0||\mathcal{N}_1\right)=\frac{1}{2}(\operatorname{tr}\left(\Sigma_1^{-1}\Sigma_0\right)+\\
+\left(\mu_1-\mu_0\right)^\top\Sigma_1^{-1}\left(\mu_1-\mu_0\right)-k+\ln\left(\frac{|\Sigma_1|}{|\Sigma_0|}\right))
\end{multline}
where $k$ is the dimensionality of the Gaussians, $\mu_i$ and $\Sigma_i$ are the mean and covariance of Gaussian $\mathcal{N}_i$, respectively, $\operatorname{tr}\left(\cdot\right)$ is the trace operator, and $|\cdot|$ is the determinant operator.

\section{Experiments}
\label{sec:experiments}

We evaluate summarization by modeling the full and summarized versions of songs with full-covariance \acp{SGM} and by computing the \ac{KL} divergence from each original \ac{SGM} to each of the corresponding summary \acp{SGM}. Since the summarizers perform well at selecting relevant and diverse information (as has been shown in the context of music classification by \cite{Raposo2015,Raposo2016}, then we should also be able to observe that performance from an information-theoretic perspective, by measuring relative entropy from original to summary \acp{SGM}. We also validate this evaluation by measuring its correlation with the evaluation of proxy tasks (i.e., genre classification and emotion regression). As baselines, we use naive summarization heuristics, namely, the beginning, middle, and end sections of the songs, as well as \ac{AvS}. These are common practice in \ac{MIR} tasks, specifically, in \ac{MIREX}, where 30s segments are considered. When evaluating the performance of the Gaussian sampler, every other of the previously mentioned summarizers and heuristics are considered as baselines to be compared against.

Note that there are two different feature extraction steps. The first is done by the summarizers, every time a song is summarized. The summarizers output the audio signal (WAV file) corresponding to the selected parts, to be used in the second step, i.e., when estimating the \acp{SGM}, classifying genre, or regressing emotion, where other sets of features are extracted from the original and summarized songs. Genre classification is done using \acp{SVM}, while emotion regression is done using \acp{SVR}, both implemented by LIBSVM \citep{Chang2011}.

\subsection{Audio Features}
\label{sub:features}

In line with \cite{Raposo2016}, we compute the first 20 \acp{MFCC} \citep{Davis1980} after a 0.05s, 0.10s, or 0.50s framing of the input signal (no overlap) as summarization features.

We evaluate information loss in a generic way by estimating the \acp{SGM} according to 2 different low-level audio descriptions: raw samples and 26-band Mel-scaled frequency spectra \citep{Stevens1937} extracted from each 0.05s.

When classifying genre, we use a 38-D vector per song, concatenating several state-of-the-art handcrafted features used in several research efforts \citep{Tzanetakis1999,deLeon2013,Foleiss2016}. These features describe the timbral texture of a music piece, consisting of the mean of the first 20 \acp{MFCC} as well as the mean and variance of 9 spectral features: centroid, spread, skewness, kurtosis, flux, rolloff, brightness, entropy, and flatness. These are computed for every non-overlapped 0.05s. We use OpenSMILE \citep{Eyben2013} for extracting this set of features.

For emotion regression, we use a 162-D vector consisting of the means and standard deviations of frame-based energy features: 10-coefficient audio power (1024-sample frames), total loudness, and 10-D \ac{SONE} \citep{Pampalk2002} (256-sample frames with 50\% overlap); temporal features (MPEG-7 standard): zero crossing rate (0.001s frames with 90\% overlap), temporal centroid, and log-attack time (0.05s frames, no overlap); spectral features (0.05s frames, 50\% overlap): centroid, spread, skewness, kurtosis, flatness, entropy, brightness, rolloff, roughness, regularity, 13 \acp{MFCC}, and inharmonicity (0.01s frames, 90\% overlap); and harmony features (0.2s frames with 95\% overlap): tonal centroid (6 chromatic-scale chord projections) and harmonic change \citep{Harte2006}, key strength (12 major and 12 minor), key clarity, and mode \citep{Gomez2006}. These handcrafted features have been widely used for \ac{MER} \citep{Yang2011}. We use \ac{MIR} Toolbox \citep{Lartillot2007} and \ac{SDT} \citep{Pampalk2004} for extracting these features.

\subsection{Datasets}
\label{sub:datasets}

The first dataset is a 1250-song 5-genre (250 songs per genre) dataset previously used by \cite{Raposo2016}. This dataset consists of full songs (47.94 to 720.06 seconds, average 282.44 seconds) from the genres of Bass, Fado, Hip-hop, Indie Rock, and Trance, and hereafter we refer to it as $G$. For this specific dataset, we trim silent segments from the beginning and end of each song as a preprocessing step in order to remove irrelevant information. Machine-oriented summarizers are specifically evaluated on this dataset according to the accuracy obtained when classifying using the corresponding summaries, in addition to the generic evaluation using relative entropy.

The second dataset is a film soundtracks dataset \citep{Eerola2011} consisting of 360 short continuous segments (10.03 to 37.02 seconds, average 17.35 seconds) of songs annotated with real values describing the emotion dimensions of arousal, valence, and tension \citep{Russel1980,Thayer1989}. Hereafter, we refer to it as $E$. Machine-oriented summarizers are specifically evaluated on this dataset according to the regression \ac{MSE} and $R^2$ correlation obtained when doing regression using the corresponding summaries, in addition to the generic evaluation using relative entropy.

\subsection{Setup}
\label{sub:setup}

We first compute each summarized version of the original datasets, converted to mono and downsampled to 22050Hz. This translates into summarizing the dataset, each song at a time, for each algorithm/heuristic and parameter combination we consider. For $G$, we consider summary durations from 5s to 30s (every 5s), whereas for $E$, we consider summary durations from 1s to 6s (every 1s), since the original signals are shorter. For the beginning, middle, and end sections, we also experimented with summary durations from 35s to 120s (every 5s) in $G$. Furthermore, we summarize the datasets with \ac{GRASSHOPPER}, LexRank, \ac{LSA}, and Support Sets, for vocabulary sizes ranging from 5\% to 30\% of the duration of the songs (in frames). We also consider 10-term phrases and use dampened tf-idf term weighting except for \ac{LSA}, where binary weighting is used \citep{Raposo2016}. We experimented with frame sizes of 0.05s, 0.10s, and 0.50s. When summarizing $E$ using the phrase-based methods, we exclude the 0.50s frame size experiments since it will lead to the original music files being described by only 2 phrases in many songs (because they are shorter than 15s), which prevents those algorithms from operating normally. For the Gaussian sampler, we experimented both with and without the mean shift correction heuristic.

After summarizing, we estimated full-covariance \acp{SGM} for all summaries, for both the raw samples and Mel-scaled spectra audio descriptions. Then, we evaluated how much information is lost by each summary, when compared to the full song, by computing its relative entropy, i.e., the \ac{KL} divergence from the original \ac{SGM} to the corresponding summary \ac{SGM} and take the average of these values for each summarized dataset version. This consists of the generic, task-agnostic evaluation procedure.

Furthermore, we also evaluated each summarized $G$ version according to the classification accuracy and evaluated each summarized $E$ version according to the regression \ac{MSE} and $R^2$. We scaled the values of arousal, valence, and tension to fit in the range of -1 to 1. We validated the generic evaluation procedure by computing the Spearman $\rho$ correlation between the generic performance and the performances of classification and regression, for $G$ and $E$, respectively. All algorithms were implemented in C++, using Eigen \citep{Eigen} for matrix operations and Marsyas \citep{Tzanetakis1999} for synthesizing the audio of the summaries.

\section{Results}
\label{sec:results}

We present results comparing the descriptive and discriminative performances of all previously mentioned summarizers. We start by measuring the information loss on both datasets by computing the average over songs of $\operatorname{D}_{\text{KL}}\left(p||q\right)$ for each summarization setup, where $p$ is the \ac{SGM} representing the original song and $q$ is the \ac{SGM} representing the summarized song. Then, for dataset $G$ we compute the classification accuracy for each summarization setup. Note that these values are different from the ones reported by \cite{Raposo2016} because, in this case, we trimmed the silences. Finally, for dataset $E$, we perform regression on the 3 emotion dimensions and compute the average \ac{MSE} and $R^2$. We only show the baseline values and the values corresponding to the parameter combination that performed the best on average (over summary durations) for each algorithm. We use $F$ to refer to frame size and offset (in seconds), $M$ to refer to the mean-shift heuristic, and $V$ to refer to vocabulary size (ratio in relation to the total number of frames). 

\begin{figure}[htb]
\begin{center}
\includegraphics[keepaspectratio=true,width=\linewidth]{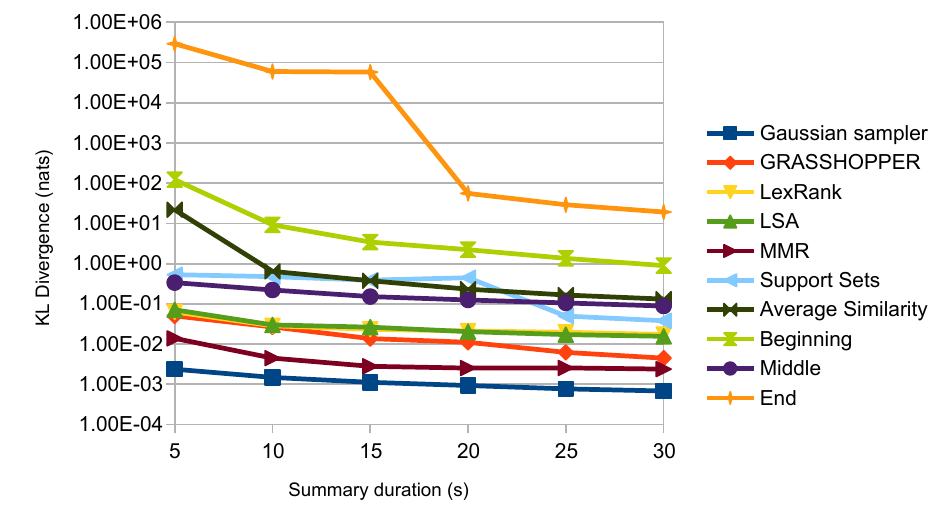}
\end{center}
\caption{Raw samples \ac{KL} divergence (in nats) -- Dataset $G$.}
\label{fig:kl-raw-classification}
\end{figure}

Figure \ref{fig:kl-raw-classification} shows the information loss (logarithmic scale) according to the distribution of raw samples on $G$. The middle section was the best continuous baseline, surpassing any of the other segment heuristics, as well as \ac{AvS} for all durations shown. Note that this is not always true, as the end sections outperform both the beginning and middle sections for durations greater than 70s (not shown in the table). Our new method (0.05 $F$, $M$) always outperforms every other approach. We also point to the fact that the mean-shift correction heuristic improves summarization performance. Every generic summarizer outperformed all baselines for all durations, with the exception of Support Sets for durations ranging from 5s to 20s, where it is outperformed by the best continuous baseline (middle sections). Note that 5s summaries produced by the Gaussian summarizer outperform all other summaries, including continuous section baseline summaries up to 120s (not shown in the table).


\begin{figure}[htb]
\begin{center}
\includegraphics[keepaspectratio=true,width=\linewidth]{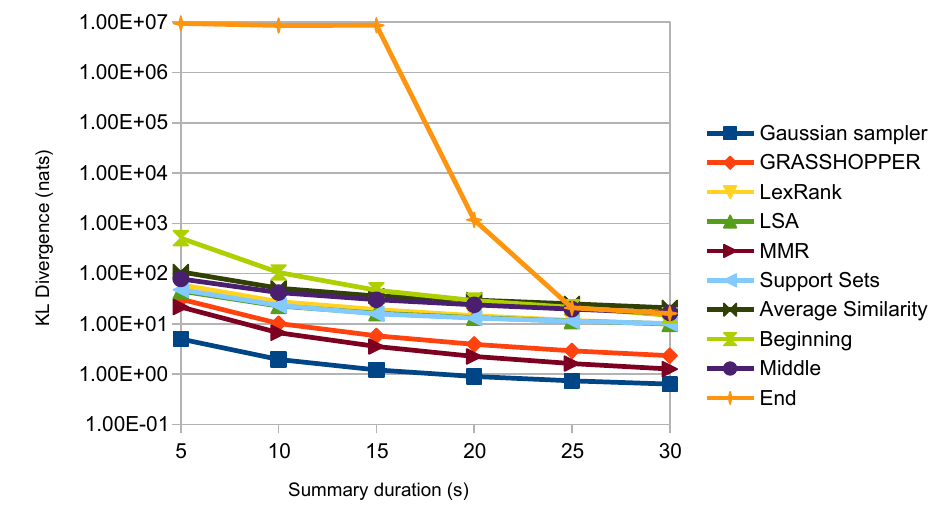}
\end{center}
\caption{Log-mel features \ac{KL} divergence (in nats) -- Dataset $G$.}
\label{fig:kl-logmel-classification}
\end{figure}

Figure \ref{fig:kl-logmel-classification} shows the information loss (logarithmic scale) according to the distribution of log-mel features on $G$. The middle section was the best continuous baseline, surpassing any of the other naive segment selection heuristics as well as \ac{AvS} in all summary durations except for 30s, where it is outperformed by the beginning and end sections. Once again, our method (0.05 $F$, $M$) outperforms every other for all summary durations using the mean-shift heuristic. Every generic summarizer outperformed all continuous baselines for all summary durations. Note that 15s summaries produced by the Gaussian sampler outperform every other, including continuous section baselines summaries up to 120s (not shown in the table).


\begin{figure}[htb]
\begin{center}
\includegraphics[keepaspectratio=true,width=\linewidth]{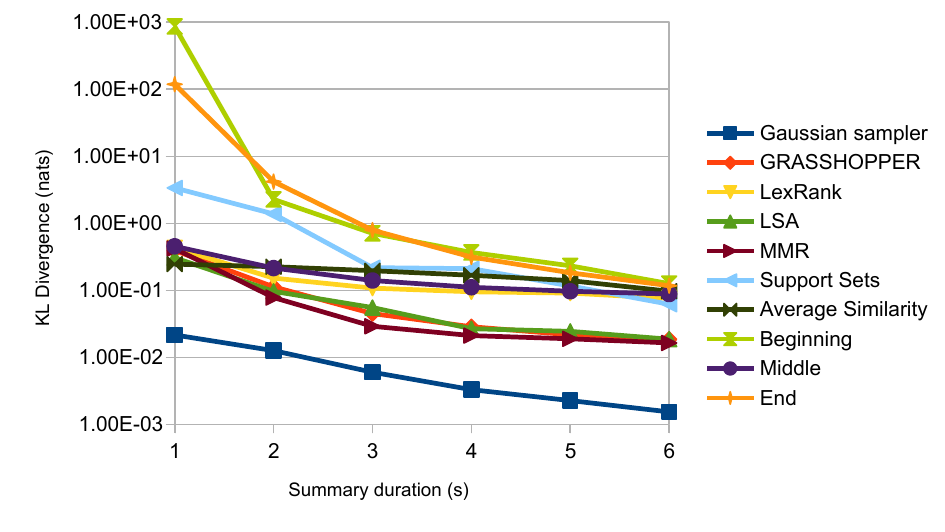}
\end{center}
\caption{Raw samples \ac{KL} divergence (in nats) -- Dataset $E$.}
\label{fig:kl-raw-regression}
\end{figure}

Figure \ref{fig:kl-raw-regression} shows the information loss (logarithmic scale) according to the distribution of raw samples on $E$. The middle section is the best continuous baseline for all summary durations, except for 1s summaries where it was outperformed by \ac{AvS}. The Gaussian sampler (0.05 $F$, $M$) always outperforms every other method. Every generic summarizer usually outperforms all continuous baselines, except for Support Sets.


\begin{figure}[htb]
\begin{center}
\includegraphics[keepaspectratio=true,width=\linewidth]{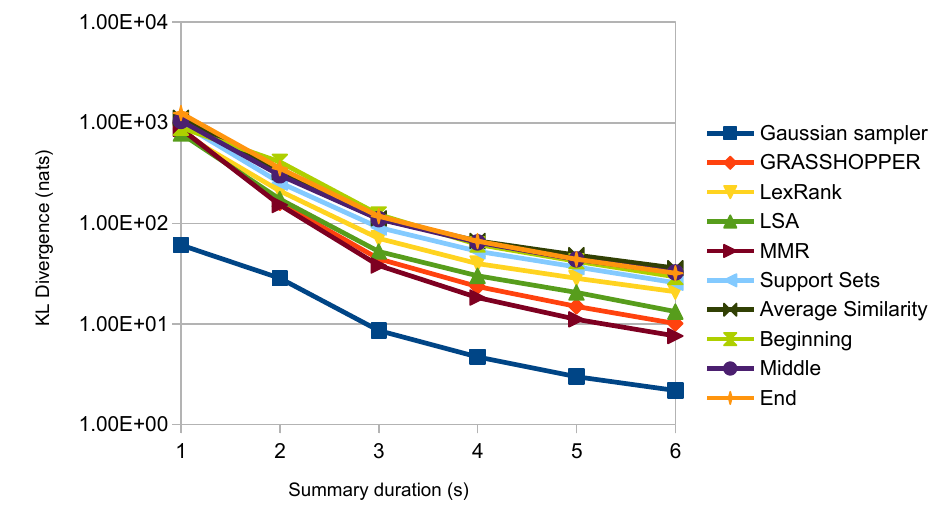}
\end{center}
\caption{Log-mel features \ac{KL} divergence (in nats) -- Dataset $E$.}
\label{fig:kl-logmel-regression}
\end{figure}

Figure \ref{fig:kl-logmel-regression} shows the information loss (logarithmic scale) according to the distribution of log-mel features on $E$. The beginning section is the best continuous baseline for all summary durations, except for 2s and 3s summaries where it is outperformed by every other summarizer. The Gaussian sampler (0.05 $F$, $M$) is still the best summarizer for all durations.


\begin{figure}[htb]
\begin{center}
\includegraphics[keepaspectratio=true,width=\linewidth]{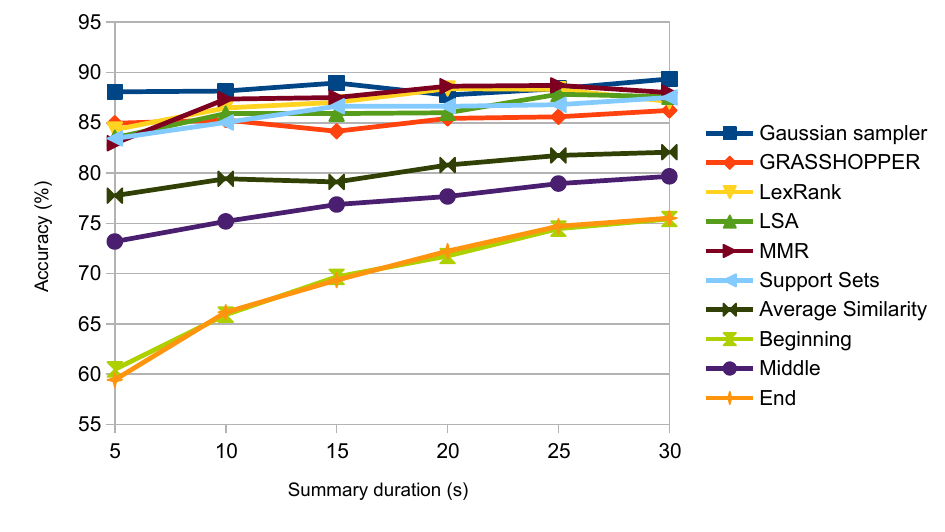}
\end{center}
\caption{Classification accuracy (\%) -- Dataset $G$ (Original dataset: 89.28\%).}
\label{fig:accuracy-classification}
\end{figure}

Figure \ref{fig:accuracy-classification} shows the music genre classification accuracy on $G$. \ac{AvS} (0.50 $F$) is the best baseline for all durations shown. The best fixed segment baseline is the middle section. However, this is not always true, as the end section baseline outperforms both the beginning and middle sections for durations greater than 40s (not shown in the table). The Gaussian sampler(0.05 $F$, $M$) is the best performing summarization method for all summary durations except for 20s and 25s. Note that 30s Gaussian sampler summaries are more discriminative than the full songs themselves and that 5s summaries are more discriminative than almost every other summaries including fixed segment baselines up to 120s, except for 110s end sections (88.96\% accuracy).

\begin{figure}[htb]
\begin{center}
\includegraphics[keepaspectratio=true,width=\linewidth]{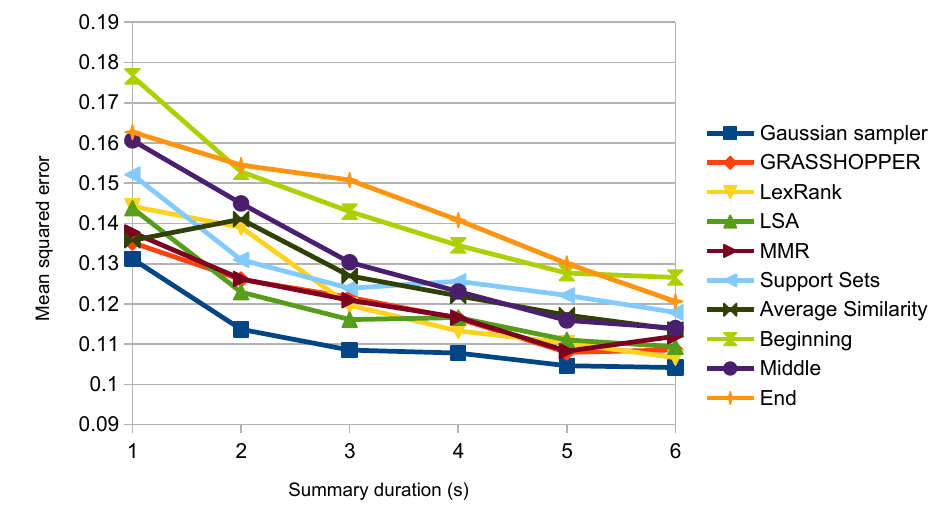}
\end{center}
\caption{Regression \ac{MSE} -- Dataset $E$ (Original dataset: 0.0966).}
\label{fig:mse-regression}
\end{figure}

\begin{figure}[htb]
\begin{center}
\includegraphics[keepaspectratio=true,width=\linewidth]{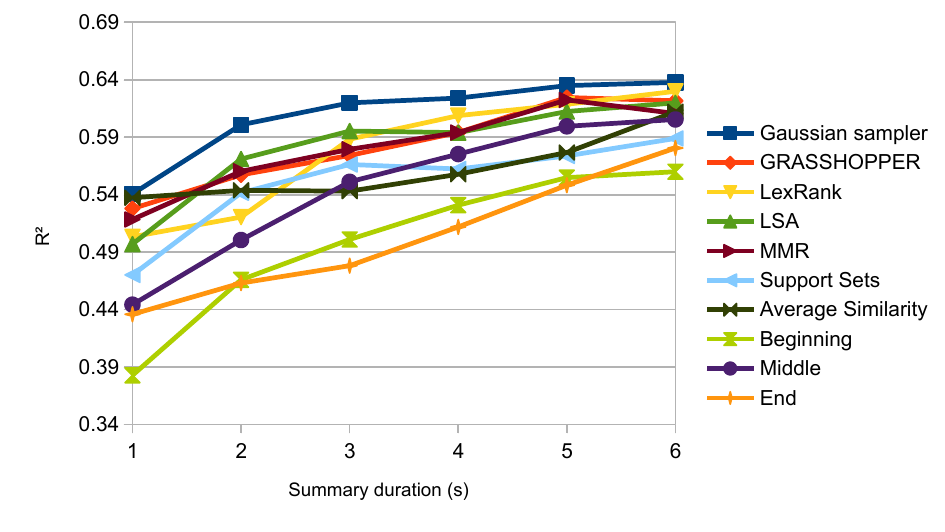}
\end{center}
\caption{Regression $R^2$ -- Dataset $E$ (Original dataset: 0.6631).}
\label{fig:r2-regression}
\end{figure}

Figures \ref{fig:mse-regression} and \ref{fig:r2-regression} show the music emotion regression \ac{MSE} and $R^2$, respectively, on $E$. \ac{AvS} (0.50 $F$ and 0.10 $F$ for \ac{MSE} and $R^2$, respectively) is the best continuous baseline, except for 5s where it is outperformed by the middle section. The Gaussian sampler (0.10 $F$) is the best for all summary durations.

\section{Discussion}
\label{sec:discussion}
Concerning fixed segment baseline results, it is intuitive to think of the middle segments as being more representative of the whole song than the beginning and end segments, since the beginning and end of songs usually differ from the rest of the song. However, even though we can observe this in both $G$ and $E$, this does not hold for longer summary durations. For longer summary durations (e.g. 70s, as observed in $G$), the beginning and end already contain much of the information that is repeated in the middle, thus becoming more representative of the whole song. Meanwhile, the middle sections just accumulate more redundant information that is present in the middle of the song. This is why, at that point, an increase in summary duration is more beneficial to the beginning and end sections.

The \ac{KL} divergence results consistently show that generic summarizers lose less information compared to the continuous baselines, which is in line with previous results \citep{Raposo2015,Raposo2016}. Moreover, they also consistently show that the Gaussian sampler outperforms every other summarization method. One could argue that evaluation through relative entropy is biased towards methods that aim at minimizing it. However, we also evaluated summarization through proxy tasks (i.e., classification and regression). The results of both proxy tasks also confirm the superiority of the Gaussian sampler.

When doing classification, the Gaussian sampler is outperformed only on 20s and 25s summaries. This is, however, due to the fact that the classification performance is almost saturated, i.e., it is very close to the performance of full songs. When we reach a certain summary duration, the difference between different summarizers becomes less noticeable when measuring it through proxy tasks (Table \ref{tab:difference}). After all, there is only so much a specific set of features and classifier can do to separate the classes. This saturation phenomenon also happens with regression albeit less noticeably. This is probably because of the smoother nature of the task results: averaged \ac{MSE} and $R^2$ of real-valued predictions of 3 emotion dimensions instead of accuracy on the hard separation between 5 classes. However, in Table \ref{tab:difference}, we can still see that the average difference between the Gaussian sampler performance and of other summarizers is lessened as the duration increases. This means that, even though it is noteworthy that the Gaussian sampler 30s summaries outperform the full songs (with less than one ninth of the duration), the real strength of this method is how short its summaries need to be in order to be close to the saturation level. The classification experiments show that 5s summaries (which is less than 56 times the duration of the original songs) achieve 88.08\% accuracy which is greater than, for instance, any 30s summaries from other methods. In fact, a Wilcoxon signed-rank test on the confusion matrices that resulted from this 5s summaries classification and full songs classification revealed that there is no statistically significant difference of using these 5s summaries from using full songs (p-value 0.1411).




\begin{table}[htb]
\footnotesize
\centering
\caption{Average performance difference}
\begin{tabular}{c|c|c|c|c|c|c}
Metric & 1s/5s & 2s/10s & 3s/15s & 4s/20s & 5s/25s & 6s/30s\\
\hline
Acc. & 4.24 & 2.14 & 2.70 & 1.34 & 1.07 & 2.06\\
\ac{MSE} & 0.0114 & 0.0153 & 0.0119 & 0.0099 & 0.0073 & 0.0067\\
$R^2$ & 0.0373 & 0.0507 & 0.0392 & 0.0334 & 0.0245 & 0.0235\\
\end{tabular}
\label{tab:difference}
\end{table}

We showed that the Gaussian sampler loses less information than all baselines, according to relative entropy. We also showed that the Gaussian sampler outperforms all baselines, according to two different proxy tasks using datasets with very different characteristics. Furthermore, we claim that relative entropy is a good and generic measure (since it does not rely on proxy tasks) of summarization performance. In order to numerically verify that claim, we computed the Spearman $\rho$ correlation between each of relative entropy results and each of the proxy task results, taking into account all different summarization setups (i.e., 433 dataset versions in $G$, and 253 dataset versions in $E$). There is a strong correlation between the performance measured by relative entropy (raw samples and log-mel spectra) and each proxy task evaluation: lower information loss means higher classification accuracy (-0.797 and -0.719), lower regression \ac{MSE} (0.852 and 0.908), and higher regression $R^2$ (-0.845 and -0.907). These values clearly suggest that relative entropy is a good generic predictor of summarization performance in the context of tasks that rely on a bag-of-features representation of objects. Intuitively, this makes sense, since representations based on bag-of-features consist necessarily of statistical descriptors of the feature distributions. Therefore, a summary whose probability distribution of features is as close as possible to the probability distribution of the features of the original object will generate similar statistical descriptors to those generated from the original object. Moreover, the fact that a summarizer aiming at minimizing relative entropy between the original and summarized data achieves state-of-the-art performance, even when evaluated by proxy tasks, further strengthens the claim that this information-theoretic way of measuring summary content is relevant for this type of tasks.


Relative entropy can measure the amount of information loss incurred by a summarizer. Thus, a summary providing a good statistical description of the original data will likely be very useful for any bag-of-features-based discriminative task, since the performance of these tasks is based on frame-level feature statistics. This work demonstrates this in two complementary ways: it shows there is a correlation between the performance in proxy tasks and information loss measured as accuracy/\ac{MSE}/$R^2$ and relative entropy, respectively; and it shows that our proposed method, which minimizes information loss, achieves state-of-the-art performance, according to all evaluations performed, in machine-oriented summarization. Furthermore, relative entropy seems to reveal subtle differences in summarization performance which are blurred out by the evaluation of proxy tasks (due to the saturation phenomenon), thereby having another advantage over proxy evaluations.

Our Gaussian sampler summarizer has several advantages for machine-oriented summarization against other generic summarizers. As opposed to previous state-of-the-art summarizers, that have several parameters to tune, the Gaussian sampler only has to consider framing (also shared by the others). This means that we no longer have the issue of finding a good vocabulary size for discretizing the frames, since no discretization is necessary. Furthermore, the smaller the frames, the longer it takes to compute the vocabulary, which makes it prohibitively expensive to compute (e.g., we found it would take too much time to try 0.05s framing for these methods, even in an experimental setting). Fine-tuning is also a problem for phrase size and, more importantly, term weighting, which is an aspect that some generic summarizers are very sensitive to \citep{Raposo2016}. Moreover, the benefits of machine-oriented summarization range from faster processing to more efficient use of bandwidth but they also include alleviating copyright issues: since the whole song is not present and the summary clip is not a continuous segment, it will not serve as enjoyment for potential listeners. Informal listening revealed that summaries with small frame sizes can even be perceived as noise. This means that \ac{MIR} datasets can be more easily distributed among the research community. The Gaussian sampler summarizer is much better in making the clip not enjoyable. This is because it picks one frame at a time resulting in many more discontinuities, as opposed to summarizers that pick whole phrases of frames.

\section{Conclusions and future work}
\label{sec:conclusions}
We generically evaluated machine-oriented music summarization in the context of bag-of-features-based tasks. Our contribution is two-fold: (1) we validated the use of the relative entropy as a generic summarization evaluation measure, by showing it is a good predictor of task-specific evaluation measures, for both music genre classification and music emotion regression; and (2) we proposed a novel and simple method for machine-oriented summarization that leverages (1) by using relative entropy from the original song to the summarized song as a criterion to be minimized. Our new method, while simple, is powerful and objectively outperforms previous state-of-the-art methods as well as facilitates dataset sharing within the \ac{MIR} community by reducing the risks of copyright infringement.

Future work includes studying topic modeling and summarization to better understand the semantics of both as well as tweaking these algorithms for human-oriented summarization.

\section*{Acknowledgments}

This work was supported by national funds through Fundação para a Ciência e a Tecnologia (FCT) with reference UID/CEC/50021/2013.

\bibliographystyle{model2-names}

\end{document}